\documentclass{elsart}

\usepackage{graphicx}

\usepackage{amssymb}
\usepackage{amsmath}

\DeclareFontShape{OT1}{cmtt}{bx}{n}{
  <5><6><7><8><9><10><10.95><12><14.4><17.28><20.74><24.88>cmttb10}{}

\begin{document}

\begin{frontmatter}


\title{DRAGON: Monte Carlo generator of particle production from a fragmented fireball 
in ultrarelativistic nuclear collisions\thanksref{label1}}
\thanks[label1]{Supported in parts by VEGA 1/4012/07 (Slovakia), MSM~6840770039, and  
LC~07048 (Czech Republic).}
\author{Boris Tom\'a\v{s}ik}
\ead{boris.tomasik@umb.sk}
\address{Fakulta~pr\'irodn\'ych~vied, Univerzita~Mateja~Bela, Tajovsk\'eho~40, 97401~Bansk\'a~Bystrica, Slovakia}
\address{In\v{s}tit\'ut matematiky a informatiky, 
Univerzita~Mateja~Bela, Cesta~na~Amfite\'ater~1, 97401~Bansk\'a~Bystrica, Slovakia}
\address{Faculty~of~Nuclear~Science~and~Physics~Engineering, Czech~Technical~University, B\v{r}ehov\'a~7, 11519~Prague,
Czech~Republic}





\begin{abstract}
A Monte Carlo generator of the final state of hadrons emitted from an ultrarelativistic
nuclear collision is introduced. An important feature of the generator is a possible fragmentation 
of the fireball and emission of the hadrons from fragments. Phase space distribution of the fragments 
is based on the blast wave model extended to azimuthally non-symmetric fireballs. 
Parameters of the model can be tuned and this allows to generate final states from various 
kinds of fireballs. A facultative output in the OSCAR1999A format allows for a comprehensive
analysis of phase-space distributions and/or use as an input for an afterburner.   
\end{abstract}

\end{frontmatter}

\section*{Program summary}

\noindent
\textit{Program title:} DRAGON\\
\textit{Catalogue identifier:}\\
\textit{Program summary URL:} http://www.fpv.umb.sk/{\symbol{126}}tomasik/dragon\\
\textit{Program obtainable from:} http://www.fpv.umb.sk/\symbol{126}tomasik/dragon\\
\textit{RAM required to execute with typical data:} 100~Mbytes \\
\textit{Number of processors used:} 1\\
\textit{Computer(s) for which the program has been designed:} PC Pentium 4, though no 
particular tuning for this machine was performed.\\
\textit{Operating system(s) for which the program has been designed:} Linux;  the program 
has been successfully run on Gentoo Linux 2.6, RedHat Linux 9, Debian Linux 4.0, all with g++
compiler. It also ran successfully on MS Windows under Microsoft 
Visual C++ 2008 Express Edition as well as under cygwin/g++.\\
\textit{Programming language:} C++\\
\textit{Size of the package:} 32~818 bytes\\
\textit{Distribution format:}  tarred and gzipped archive   \\
\textit{Number of lines in distributed program, including test data etc.:} 6368    \\
\textit{Number of bytes in distributed program, including test data etc.:} 153~939   \\
\textit{Nature of physical problem:}
Deconfined matter produced in ultrarelativistic nuclear collisions expands and cools down
and eventually returns into the confined phase. If the expansion is fast, the fireball could fragment 
either due to spinodal decomposition or due to suddenly arising bulk viscous force. Particle 
abundances  are
reasonably well described with just a few parameters within the statistical approach.
Momentum spectra integrated over many events can be interpreted as  produced from 
an expanding and locally thermalised fireball. The present Monte Carlo model unifies
these approaches: fireball decays into fragments of some characteristic size. The fragments
recede from each other as given by the pre-existing expansion of the fireball. They 
subsequently emit stable and unstable hadrons with momenta generated according to thermal 
distribution. Resonances then decay and their daughters acquire 
momenta as dictated by decay kinematics. 
\\
\textit{Method of solving the problem:}
The Monte Carlo generator repeats a loop in which it generates individual events. 
First, sizes of fragments are generated. Then the fragments are placed
within the decaying fireball and their velocities are determined from the one-to-one 
correspondence between the position and the expansion velocity in the blast wave model. 
Since hadrons may be emitted from fragments as well as from the remaining bulk fireball, 
first those from the bulk are generated according to the blast wave model. Then, hadron
production from the fragments is treated. Each hadron is generated in the rest frame of the 
fragment and then boosted to the global frame. Finally, after all directly produced hadrons 
are generated, resonance decay channels are chosen and the momenta and  
positions of final state hadrons are determined. 
\\
\textit{Typical running time:} Generation of 10$^4$ events can take anything between 2 hours to a couple of days. 
This depends mainly on the size and density of fragments. Simulations with small fragments may be very slow.  
At the beginning of a run there is a period of up to 1 hour in which the program calculates thermal weights 
due to statistical model. This period is long if many species are included in the simulation.

\section{Introduction}
\label{intro}

In ultrarelativistic collisions of heavy atomic nuclei matter is probed 
at high temperature and density. The fireball thus created exists only 
for very short period of time; it quickly expands and hadrons decouple 
from it. If the collision energy is high enough, a phase of deconfined 
quarks and gluons is reached in the early phase of the collision. Due to 
pre-existing longitudinal movement of the incident nucleons and the inner
pressure of the matter, it swiftly expands in longitudinal as well as
transverse direction. At certain moment the energy density becomes too low
to justify deconfinement and the matter hadronizes. From lattice QCD calculations
we know that the change from deconfined to confined matter is  rapid though 
smooth crossover in the region of the phase diagram at low baryochemical potential 
\cite{Aoki:2006we} and it becomes first order phase transition from certain value of
$\mu_B$ upwards \cite{Fodor:2004nz,deForcrand:2003hx,Allton:2003vx}.

It is important for our discussion that the passage through this transition 
is very fast. In such a case, equilibrium description may be inapplicable and 
the phase transition would not proceed in the same way as probed on lattice. 
Instead, considerable supercooling can occur. In the region of the phase diagram 
where first order phase transition is expected even spinodal may be reached 
by the system if the expansion rate is bigger than the nucleation rate of bubbles
of the new phase \cite{Mishustin:1998eq,Scavenius:2000bb}. Then, the fireball disintegrates into droplets
similarly to spinodal fragmentation known in liquid/gas nuclear phase 
transition \cite{Chomaz:2003dz}. 

Seemingly, such a mechanism should not work in the region of the phase diagram with 
smooth crossover. Nevertheless, it has been argued that an abrupt rise of 
bulk viscosity at $T_c$ can suddenly make the fireball very stiff and if 
strong expansion is present in such a moment it can drive the system into 
fragmentation \cite{Torrieri:2007fb}. Hydrodynamic expansion in nuclear collisions
is possibly unstable \cite{Torrieri:2008ip}, thus fragmentation at hadronisation phase 
transition seems likely scenario. 

On the other hand, hydrodynamically inspired parametrisations 
\cite{Csorgo:1995bi,Csanad:2003qa,Tomasik:1999cq,Retiere:2003kf,Broniowski:2001we}
combined with thermal models 
\cite{Torrieri:2004zz,Cleymans:2005xv,BraunMunzinger:2001ip,Becattini:2003wp} provide
satisfying description of single-particle spectra, particle abundances, and 
some also femtoscopy measurements. It is necessary to note that these are 
observables which are extracted from data {\em summed} over a large number 
of events. Event-by-event fluctuations of mean $p_t$ \cite{Adcox:2002pa,Adams:2003uw,Broniowski:2005ae} 
and angular correlations \cite{Li:2006zzd} indicate possible presence of clusters 
in momentum distributions. Such clusters could be due to fragmentation at the 
phase transition. Note that momentum clusters are buried under many entries 
to the histograms if summed over many events. 

It is the purpose of the present Monte Carlo droplet generator to produce
artificial data sets which resemble those coming from real nuclear collisions
provided fragmentation occurs at hadronisation and hadrons are emitted from 
fragments without any further scattering. 
Its name, DRAGON, stands for DRoplet and hAdron GeneratOr for Nuclear collisions. 
In a way, the model is similar to 
THERMINATOR \cite{therminator}, with the crucial difference that emission 
from fragments is included. Note that the code can write out the 
final state in OSCAR1999A format \cite{oscar} and thus a possible further evolution
of the hadronic cloud can be simulated with the help of a cascade generator. 

In the next Section the model of particle emission from a fragmented fireball 
is reviewed. Section \ref{s:struct} explains the architecture of the program. 
Sections \ref{s:installing} and \ref{s:running} explain how to install and run 
the generator. Some representative results are presented in Section \ref{s:results}
and the paper in concluded in Section \ref{s:conc}. Details about generation 
of momenta from Boltzmann distribution are summarised in the Appendix.


\section{The model of particle emission from fragmented fireball}

In rapid phase changes the fireball can fragment and hadrons are emitted 
from the produced  fragments. There may also be a portion of the produced hadrons which 
is emitted directly from the bulk fireball. 

\subsection{Hadrons produced from bulk}

Directly produced hadrons are described by the \emph{blast-wave model}. The Wigner 
distribution of their emission points and momenta is given as \cite{Retiere:2003kf,tomasik_app}
\begin{multline}
\label{emisf}
S(x,p)\, d^4x = \frac{2s+1}{(2\pi)^3}\,m_t \, \cosh(y-\eta)\,
\exp\left( - \frac{p^\mu u_\mu}{T_k}\right )\, \\
\times\Theta(1-\tilde r(r,\phi))\, H(\eta) 
     \,\delta(\tau - \tau_0)d\tau\, \tau\, d\eta\, r\, dr\, d\phi\, .
\end{multline}
The model is formulated in terms of relativistic and polar coordinates $r$, $\phi$, $\eta$, and 
$\tau$
\begin{subequations}
\begin{eqnarray}
x^0 & = & \tau \cosh\eta\\
x^1 & = & r\cos\phi\\
x^2 & = & r\sin\phi\\
x^3 & = & \tau \sinh\eta\, ,
\end{eqnarray}
\end{subequations}
while for momentum we use rapidity $y$, transverse mass (momentum) $m_t$ ($p_t$), and 
azimuthal angle $\psi$
\begin{subequations}
\begin{eqnarray}
p^0 & = & m_t \cosh y\\
p^1 & = & p_t \cos \psi\\
p^2 & = & p_t \sin \psi\\
p^3 & = & m_t \sinh y\, . 
\end{eqnarray}
\end{subequations}
Emission points are distributed uniformly in radial direction for 
\begin{equation}
\tilde r = \sqrt{\frac{(x^1)^2}{R_x^2} + \frac{(x^2)^2}{R_y^2}} < 1\, ,
\end{equation}
where 
\begin{equation}
\label{rxry}
R_x = a\, R \, , \qquad R_y = \frac{R}{a}\, ,
\end{equation}
with $R$ being the mean transverse radius of the ellipsoidal fireball and $a$ its 
spatial deformation parameter. 

The distribution $H(\eta)$ specifies the profile of the fireball in space-time  
rapidity. Wigner density in eq.\eqref{emisf} is written in Boltzmann approximation and the factor 
$p_\mu u^\mu$ gives the energy of the produced hadron in the rest frame of the moving 
fluid. The fluid velocity field is parametrised as
\begin{eqnarray}
\label{velo}
u_{\mu}  & = & 
(\cosh\!\eta\, \cosh\!\eta_t,\, \cos\!\phi_b\, \sinh\!\eta_t,\,
\sin\!\phi_b\, \sinh\!\eta_t,\, \sinh\!\eta\, \cosh\!\eta_t)\, , \\
\tan\phi_b & = & a^4 \tan\phi\, ,\\
\label{etat}
\eta_t  & = & \tilde r \, \rho_0\, \sqrt{2}\left (1 + \rho_2 \cos(2\phi_b)\right )\, .
\end{eqnarray}
Finally, the factor $(2s+1)$ in eq.~\eqref{emisf} stands for the degeneration due to 
spin. 

Formulated in this way, the fireball can have elliptic transverse shape
controlled by the parameter $a$ and elliptic transverse flow profile parameterized by 
$\rho_2$.

\subsection{Hadrons emitted by fragments}
\label{ss:hfrag}

The fireball decays into fragments of spherical shape (in their 
rest frame). They are placed according to the  distribution \eqref{emisf} 
with $T \to  0$, i.e.\ their velocity is identical the local fluid velocity at the 
place at which they were produced.

Fragments may come in one given volume $b$, if they are produced by mechanism
which leads to one length scale which dominates the fragmentation---like spinodal 
fragmentation. Another possibility, implemented in the model is that the volumes 
are distributed according to gamma-distribution
\begin{equation}
\label{gammadist}
{\cal P}_k(V) = \frac{1}{b\Gamma(k)}\, \left ( \frac{V}{b}\right )^{k-1}\, \exp\left (-\frac{V}{b}\right )\, ,
\end{equation}
with $k=2$ \cite{Mishustin:2005zt}\footnote{%
Note the different use of the parameter $b$ here and in \cite{Mishustin:2005zt}: while there
it has the dimension of inverse volume, here it has the dimension of volume. This brings it here
on equal footing with the case when all fragments have the same volume.}.

Fragments decay into hadrons exponentially in time, so the distribution of the emission 
time of a hadron in the rest frame of a fragment is 
\begin{equation}
\label{expdec}
{\cal P}_t(\tau_d) =  \frac{1}{R_d} \exp\left ( - \frac{\tau_d}{R_d} \right )
\end{equation}
where $R_d$ is the radius of the droplet (fragment). Hadrons are produced from the
whole volume of the fragment with uniform probability. Their momentum is chosen according 
to Boltzmann distribution with the temperature $T_k$ in the rest frame of the fragment.

\subsection{Resonance decays}
\label{reso}

Resonances may be produced from the fireball. Their lifetime is random 
according to exponential decay law $\exp(- \Gamma \tau)$ (in the rest frame 
of the resonance). 

If a resonance with mass $M$ decays via two-body decay to daughters with 
masses $m_1$ and $m_2$, their energies will be 
\begin{eqnarray}
E_1 & = & \frac{M^2 - m_2^2 + m_1^2}{2M} \\
E_2 & = & \frac{M^2 - m_1^2 + m_2^2}{2M}
\end{eqnarray}
and they will be receding back-to-back (in the rest-frame of the resonance) with 
momenta
\begin{equation}
\left | \vec p_1\right | = \left | \vec p_2 \right | = 
\frac{\sqrt{\left ( M^2 - (m_1 + m_2)^2   \right )\left ( M^2 - (m_1 - m_2)^2  \right )   }}{2M}\, .
\end{equation}

In case of three-body decays, in the rest-frame of the resonance, all the  momenta of the daughter 
particles lay within a plane. Under the assumption of transition amplitude independent 
of momenta, the energy distributions of daughter particles are uniform. Thus there is some freedom
in choosing the energies and momenta of daughter particles in such a way that the energy and momentum 
are conserved
\begin{subequations}
\begin{eqnarray}
E_1 + E_2 + E_3 & = & M\\
\left | \vec p_1\right |^2 + \left | \vec p_2 \right |^2 + 2\left | \vec p_1\right | \left | \vec p_2 \right |
\cos{\theta_{12}} & = & p_3
\end{eqnarray}
\end{subequations}
where $E_i = \sqrt{\left| \vec p_i\right |^2 + m_i^2}$ and $\theta_{12}$ is the angle between the momenta
$\vec p_1$ and $\vec p_2$.

Daughter particles from a resonance decay may also be unstable. In that case, they will decay according 
to the same procedure.


\subsection{Chemical composition}

Relative abundances of the individual species follow the prescription of  
chemical equilibrium with temperature $T_{ch}$ and chemical potentials for baryon number 
and strangeness $\mu_B$, and $\mu_S$. The density of species $i$ is then given as
\begin{multline}
n_i(T_{ch},\mu_b,\mu_S) = g_i \int \frac{d^3p}{(2\pi)^3} \, 
\left [ \exp\left ( 
\frac{\sqrt{p^2 + m_i^2} - (\mu_BB_i + \mu_SS_i)}{T_{ch}}\right ) \mp 1\right ]^{-1} \\ 
= \frac{g_i}{2\pi^2} T_{ch}^3\, 
I \left ( \frac{m_i}{T_{ch}},\, \frac{\mu_i}{T_{ch}} \right )\, ,
\end{multline}
where the upper (lower) sign is for bosons (fermions), $g_i$ is the degeneracy factor, and 
\begin{eqnarray}
I \left ( \frac{m_i}{T_{ch}},\, \frac{\mu_i}{T_{ch}} \right ) & = & \int_0^\infty dx\, x^2 \,
\left [ \exp \left ( \sqrt{x^2 + \frac{m_i^2}{T_{ch}^2}} - 
\frac{\mu_SS_i + \mu_BB_i}{T_{ch}} \right )
\mp 1 \right ]^{-1} \\
\mu_i  & = & \mu_B B_i + \mu_S S_i\, .
\end{eqnarray}
From this, the probability that a random particle belongs to species $i$ is 
\begin{equation}
\label{weights}
w_i(T_{ch},\mu_B,\mu_S) = \frac{n_i(T_{ch},\mu_B,\mu_S)}{\sum_i n_i(T_{ch},\mu_B,\mu_S)}\, ,
\end{equation}
where the sum in the denominator runs through all the species.


\section{Programming structure and solution}
\label{s:struct}

\begin{figure}
\centerline{\includegraphics[height=19cm]{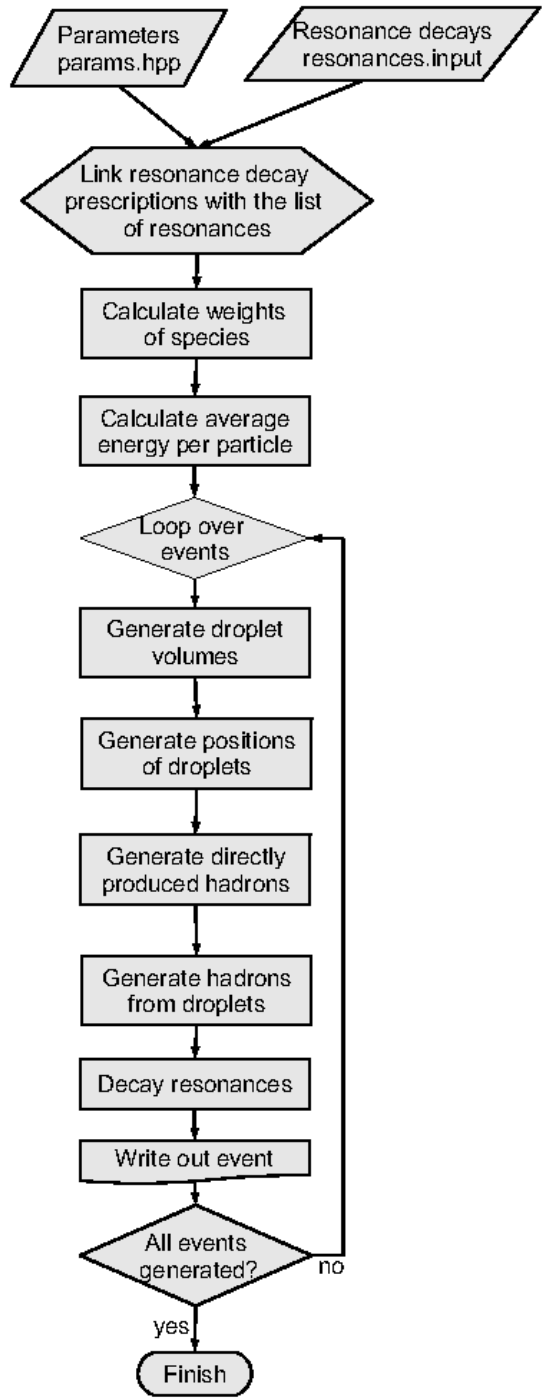}}
\caption{\label{f:structure} 
The structure of the program.}
\end{figure}

The structure of the program is outlined in Figure~\ref{f:structure}.  It 
is written in C++.
%
%
The code is split into three files: \texttt{dropem.cpp} includes the main 
function and functions for reading and initialising resonance decays and
for their decay. The file \texttt{dgener.cpp} defines classes needed for 
the working of the MC generator, and \texttt{specrel.cpp} defines 
basic vectors and tensors and operations with them.

\subsection{Introductory phase}

Parameters of the model and steering constants for compiling and running 
are specified in the file \texttt{params.hpp}. This also includes the list of 
all species called \texttt{pproperties}. This is an array, whose entries 
are records for individual species. 
In the current version we standardly include all baryons with masses 
up to 2.0~GeV/$c^2$ and mesons up to 1.5~GeV/$c^2$. 

Resonance decays are listed in a separate input file 
\texttt{resonances.input}. A file with this name must be present in the 
directory in which the program is running. It lists the modes in which resonances 
can decay. If at least one decay mode for given species is listed in this file, that species is 
treated as an unstable resonance and, vice versa, if species is not listed in the 
resonance file, it is stable. In order to store data of all decay modes of a 
resonance and simulate the decay of resonances, a class \texttt{DecayPattern}
is defined in file \texttt{dgener.cpp}. After all decay prescriptions are
read in,   a link to the corresponding \texttt{DecayPattern} is added to 
each record of unstable species in \texttt{pproperties}. Vice versa, 
daughter particle species defined in the decay modes are linked to their properties
in \texttt{pproperties}. 

Subsequently, weights for generation of all species are calculated according to 
eq.~\eqref{weights} and stored for each sort of hadrons in the list 
\texttt{pproperties}.

At the end of the initial phase,  average energy of  hadron at the specified chemical 
composition and kinetic temperature is determined. This allows to translate the expected 
multiplicity (specified as input parameter) into the energy which is contained in the fireball 
or the fragments. With the help of the energy density this is translated into the 
expected volume of the fragments. This later controls the number of generated fragments.

\subsection{Loop over events}

The number of events to be generated is given as a macro \texttt{NOEvents} in parameter file
\texttt{params.hpp}. The loop is repeated this number of times.

DRAGON can generate fireballs with ellipsoidal cross-section, so first the direction 
of the event plane is chosen randomly. 
Thus each event has a different reaction plane as it is the case in real data.

In the next step, sizes of fragments are either generated according to gamma-distribution, 
eq.~\eqref{gammadist}, or they are set all to the same value. Their number is chosen so 
that the energy which they contain corresponds to the multiplicity to be produced from fragments. 

The sizes, if they are not all equal, are then sorted with the help of a quicksort algorithm. 
This is done in order to facilitate placing of the fragments in the reaction volume. Their positions are 
generated randomly as specified in Section~\ref{ss:hfrag}. Rapidity of the fragments is chosen 
randomly according to either Gaussian distribution with specified mean and width, or uniform 
distribution between specified minimum and maximum. Transverse position is chosen uniformly within 
the ellipsis with radii $R_x$ and $R_y$. The direction of  $R_x$ is parallel to the event plane. 
Velocity of the fragment is then specified by eq.~\eqref{velo}.
For each fragment, after its position
is generated, it is checked if it does not overlap with any of the previously allocated fragments. 
If so, the position is generated anew.
The algorithm starts with large fragments and continues to the smaller ones; otherwise there could
appear problems with placing the big fragments at the end. 

Once all fragments are placed within the fireball, hadrons emitted from the bulk (not from fragments)
are generated. This begins with determining the type of hadron according to the probabilities
calculated in the initial stage. After this, the thermal component of hadron momentum is generated
according to Boltzmann distribution with temperature $T_k$. Rejection method is used; more 
details can be found in the Appendix. The direction of the momentum is random with 
isotropic distribution. The position from which hadron is emitted is chosen according to the emission 
function in eq.~\eqref{emisf} with the same rapidity distribution as was used for fragments. The 
position corresponds to some value of the collective expansion velocity via eq.~\eqref{velo}.
The generated momentum is then boosted by this velocity. It is also checked that the generated 
particle is not produced within some of the fragments. If so, its position is rejected and generated
again. The whole procedure is repeated until the expected number of hadrons from the bulk is generated.  

Hadrons emitted from fragments are generated in a similar way as those from the bulk. 
In the rest frame of the fragment, energy and momentum 
are generated according to Boltzmann distribution and the position according to homogeneous 
distribution. The times of emission are distributed exponentially, as seen from eq.~\eqref{expdec}. 
Finally, hadron position and momentum are boosted by the velocity of the fragment and the position is  
also shifted by the initial position of the fragment. After the
generation of each hadron it is checked whether the energy contained in all hadrons from 
given fragment so far does not exceed the total energy of the fragment. If it does, no more hadrons 
are generated from that fragment. Thus energy and momentum are not strictly conserved in hadron 
emission from fragments. This is not bothering as long as 
the energy and momentum of the fragment are unknown in the experiment. 

The next step is to decay resonances. This is performed
by void function \texttt{DecayResonances}. All hadrons 
are stored in a First-In-First-Out stack of particle records. The record for each particle 
includes a link to its decay prescription; if the particle is stable then no link is present. The 
stack also keeps the number of particles. The algorithm counts stable particles. It takes one particle 
from the stack. If it is stable, the particle is put back into the stack and the counter 
of stable particles is increased by one. If it is unstable, then decay is initiated. The counter of stable particles
is reset to 0 and daughter particles are put into the stack. This is repeated until the counter of stable 
particles reaches the number of particles in the stack. 

To simulate the decay of a resonance the algorithm first chooses the decay channel according 
to branching ratios. Momenta of daughter particles in the rest frame of the fragment are determined 
as explained is Section \ref{reso}. These momenta are first rotated so that their orientation 
is distributed isotropically and then boosted by the resonance velocity. 
Then, a time in which the resonance decays is generated according to 
the exponential decay law. The position in which the resonance finds itself at that time is the 
initial position of the daughter particles. 

When all resonances are decayed, generated hadrons are written into the output file.


\section{Installing DRAGON}
\label{s:installing}

The droplet generator DRAGON is distributed as a package with three C++ header 
files (\texttt{specrel.hpp}, \texttt{dgener.hpp}, \texttt{params.hpp}), three C++
source files (\texttt{specrel.cpp}, \texttt{dgener.cpp}, \texttt{dropem.cpp}),  
an input file with resonance decays (\texttt{resonances.input}) and 
\texttt{Makefile} for easy compiling on Linux. 

\begin{figure}
\centerline{\includegraphics[width=8cm]{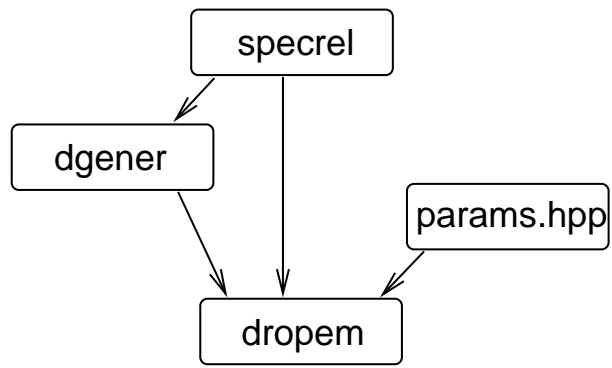}}
\caption{\label{f:dep} 
Dependences of the source files.}
\end{figure}

The dependences of individual source files in the distribution package are 
illustrated in Figure~\ref{f:dep}.
Their content is as follows:
\begin{description}
\item{\texttt{\textbf{specrel}}} contains prototypes (\texttt{.hpp}) and 
definitions (\texttt{.cpp}) of objects for four-vectors and three-vectors together
with operations on them, like addition and multiplication with Minkowski metric, where 
applicable. Also tensors and their operations are introduced. There are functions for 
boosting four-vectors to other reference frames. Also an object \texttt{Particle} is defined
which stores properties, momentum and emission position of a particle.
\item{\texttt{\textbf{dgener}}} (\texttt{.hpp} and \texttt{.cpp} file) introduces 
specific technical tools which are used in the random generator.
\item{\texttt{\textbf{dropem.cpp}}} contains the main routine with the code according to
algorithm outlined in Section~\ref{s:struct}. 
\item{\texttt{\textbf{params.hpp}}} is the header file with settings and values for parameters
which are used in the generation. This file can be edited in order to 
compile and run the generator with a different set of parameters. 
\item{\texttt{\textbf{resonances.input}}} is input data file which should be present (under this name)
in the directory in which the generator is running. 
\end{description}

The generator should compile and run on any system with standard C++ compiler. It has been tested
with  \texttt{g++} on Gentoo Linux 2.6, RedHat Linux 9, 
Debian Linux 4.0. It also ran successfully on MS Windows under Microsoft 
Visual C++ 2008 Express Edition as well as under cygwin/g++.


\section{Running DRAGON}
\label{s:running}

On Linux, \texttt{Makefile} makes sure that executing the \texttt{make} command will 
compile, link and prepare the executable file \texttt{dragon.exe} correctly. If \texttt{make} cannot be 
used, then one should follow the commands as they appear in the \texttt{Makefile}. Visual C++ by
Microsoft will compile and link the code if all \texttt{.cpp} and \texttt{.hpp} files  
are included. 

The executable \texttt{dragon.exe} is run with one or two parameters like e.g.
\begin{quote}
\texttt{
> dragon.exe outfile.out dropinfo.out
}
\end{quote}
The first parameter is the name of the output file, where generated particles will be written out by the program, 
in an output format which can be chosen. The second parameter, which is optional, stores information 
about the droplets which have been generated in each event. If the second parameter is not given, this
information is not written out. 
If parameters are omitted, then the default output file \texttt{DRAGON\_events.out} is chosen. A file 
\texttt{resonances.input} must be present in the directory where the program is being run.

\subsection{Input}

Settings for the program and parameters are chosen before the compilation in the file \texttt{params.hpp}. 
They are also explained in the comments within the file. Here, they are explained in more detail:
\begin{description}
\item{\texttt{\textbf{NOEvents}}} is the number of events which shall be generated
\item{\texttt{\textbf{RANDOMDROPLETS}}} is a logical constant; which is set either to 1, if the volumes of 
fragments are to be set randomly according to eq.~\eqref{gammadist}, or to 0 if all fragment volumes
should be equal
\item{\texttt{\textbf{OVERLAPFORBIDDEN}}} is by default set to 1; it can be set to 0 if the user 
wants to allow  an overlap of fragments. The latter choice is unrealistic, but may be useful 
in some special studies, since forbidding fragment overlap leads to some anticorrelation of 
momenta when hadrons from different droplets are usually unlikely to have similar momenta.
\item{\texttt{\textbf{GAUSSIAN\_RAPIDITY}}}
is a logical constant which is set to 1 if the rapidity profile of the fireball should be Gaussian 
and to 0 for uniform rapidity profile
\item{\texttt{\textbf{ACCEPTANCECUT}}} declares which hadrons should be written out to the output
file. This can be used to formulate  simple conditions for detector acceptance. 
The macro which is defined here is a logical statement which is evaluated for each  particle
before it is written out. Following kinematic variables can be used: rapidity (defined as variable \texttt{yrap}), 
transverse momentum (\texttt{pT}), or azimuthal angle (\texttt{Phi}). The condition can be, e.g. 
\begin{quote}
\texttt{
\#define ACCEPTANCECUT ((yrap>-1.)\&\&(yrap<=1.))
}
\end{quote}
i.e.\ hadrons with 
rapidity between -1 and 1 are recorded.
\item{\texttt{\textbf{WRITEOUT}}} is an optional string which is written in the header part of 
the output file if it is in OSCAR1999A format \cite{oscar}. By default, it is left empty. If a
newline command `$\backslash$n' appears, it should be followed by `\#', which starts a comment
line in OSCAR1999A format. 
\item{\texttt{\textbf{short FORMAT}}} is an identifier of the format of the output file. Currently, 
four possible output formats are predefined. They are explained in the comments. Note that in 
the OSCAR1999A format  two numbers in each line are added to the standard output. The last 
number is 1 if the hadron described in that line comes from resonance decay and 0 otherwise. 
The next to last number identifies the fragment from which the hadron (or its parent resonance) 
stems; and is --1 for hadrons not originating from any fragment. 
\item{\texttt{\textbf{BIGMASS}}} is set to 10.\ by default. This is the parameter $B$ described 
in the Appendix below equation~\eqref{acc-rel}. Its setting requires some fine-tuning  and 
we do not recommend to change it.
\item{\texttt{\textbf{Anr}}} is set to 20.\ by default. This is the parameter $A$ described in 
the Appendix in eq.~\eqref{acc-nonrel}. It is not recommended to change it.
\item{\texttt{\textbf{double fotemp}}} is the \emph{kinetic} freeze-out temperature in units of GeV.
\item{\texttt{\textbf{double Tch}}} is the \emph{chemical} freeze-out temperature in units of GeV.
\item{\texttt{\textbf{double mub}}} is the baryochemical potential in GeV.
\item{\texttt{\textbf{double mus}}} is the chemical potential for strangeness in GeV.
\item{\texttt{\textbf{double huen}}} is the energy density within the fragments, in units of 
GeV.fm$^{-3}$.
\item{\texttt{\textbf{double minrap}}} is the minimum rapidity of fragments or directly 
produced hadrons which will be generated.
\item{\texttt{\textbf{double maxrap}}} is the maximum rapidity of fragments or directly 
produced hadron which will be generated, so fragments and direct hadrons are generated with 
rapidities between \texttt{minrap} and \texttt{maxrap}.
\item{\texttt{\textbf{double dNdy\_total}}} is the ${dN}/{dy}$ of \emph{all} hadrons which 
roughly should be generated. This variable is useful in case of uniform rapidity distribution 
and is only used for calculation of \texttt{N\_total} (see next). If it is not used there, it 
can be commented out.
\item{\texttt{\textbf{double N\_total}}} is the total expected multiplicity in the interval between 
rapidities \texttt{minrap} and \texttt{maxrap}. For a uniform rapidity distribution it can be conveniently 
calculated as 
\begin{quote}
\texttt{
double N\_total = dNdy\_total * (maxrap-minrap);
}
\end{quote}
It can also be set equal to a given number.
\item{\texttt{\textbf{double DropletPart}}} is a parameter between 0 and 1 which determines the fraction of all hadrons that stem from the decay of fragments. It is 1 if all hadrons are produced from fragments and 
0 if they all are emitted directly from the bulk fireball. 
\item{\texttt{\textbf{double rapcenter}}} is only relevant for Gaussian rapidity distribution and gives
the rapidity of the centre of the rapidity profile. It is irrelevant if uniform rapidity distribution is 
chosen.
\item{\texttt{\textbf{double rapwidth}}} is the width of Gaussian rapidity distribution and is irrelevant if 
uniform rapidity distribution is chosen.
\item{\texttt{\textbf{double rb}}} is the radius, in fermi, of the fireball which decays into 
fragments and/or hadrons. It appears as $R$ in eq.~\eqref{rxry}.
\item{\texttt{\textbf{double a\_space}}} is the spatial anisotropy parameter $a$ as it appears in 
eq.~\eqref{rxry}. 
\item{\texttt{\textbf{double tau}}} is the parameter $\tau_0$ from eq.~\eqref{emisf}, in units of 
fm/$c$.
\item{\texttt{\textbf{double etaf}}} is the parameter $\rho_0$ from eq.~\eqref{etat}.
\item{\texttt{\textbf{double rho2}}} is the parameter $\rho_2$ from eq.~\eqref{etat}.
\item{\texttt{\textbf{double b}}} is the parameter $b$ of the gamma-distribution of fragment 
sizes in eq.~\eqref{gammadist}, if \texttt{RANDOMDROPLETS} is set to 1. In case that 
\texttt{RANDOMDROPLETS} is set to 0, this gives the volume of fragments.
\item{\texttt{\textbf{long int seed}}} is the initial seed for pseudo-random generator. If 
this is set to 0, the generator is initiated with machine time.
\item{\texttt{\textbf{int NOSpec}}} is the number of species which can be generated in the 
simulation. 
\item{\texttt{\textbf{PChem pproperties[]}}} is a vector of structures \texttt{PChem} which 
store the records of properties of individual species. The vector has as many entries as specified 
by \texttt{NOSpec}. For one species, these properties must be given (in this ordering),
\begin{enumerate}
\item Monte Carlo ID number of species according to Particle Data Group \cite{pdg}; integer
\item mass in GeV/$c^2$; double 
\item baryon number; integer
\item strangeness; integer
\item 1 if the species is boson or 0 if it is fermion
\item spin degeneracy; integer
\item put 1. here, double; (this will be calculated later by the program)
\item put another 1. here, double; (this will be calculated later by the program)
\item put --1 in the last position, integer; this will be also determined by the program and 
will link the species to its decay prescriptions; it will remain --1 for stable particles.
\end{enumerate}
Data entries for one species are divided by commas. Record for one species should be input within curly 
brackets. 
\end{description}

Decays of resonances are listed in the file \texttt{resonances.input}. The structure of the 
records in that file is as follows 
(an example of the file \texttt{resonances.input} is shown in Figure~\ref{f:resoin}):
\begin{figure}
\centerline{
\footnotesize{\texttt{
\begin{tabular}{lllllll}
\# rho+ \\[-2ex]
213	& 0.766	& 0.150\\[-2ex]
1.	& 211	& 0.13957	&   111	& 0.13498\\[-2ex]
\\[-2ex]
\# rho0\\[-2ex]
113&	0.769	&0.151\\[-2ex]
1.	&211	&0.13957	&   -211&	0.13957\\[-2ex]
\\[-2ex]
\# rho-\\[-2ex]
-213&	-1.&	0.150\\[-2ex]
1.&	-211&	-1.	&   111&	-1.\\[-2ex]
\\[-2ex]
\# omega\\[-2ex]
223&	0.782&	0.00844\\[-2ex]
0.888	&-211&	0.13957	&   111&	0.13498	&  211&	0.13957\\[-2ex]
0.0221&	-211&	0.13957	 &  211&	0.13957\\[-2ex]
0.085&	111&	0.13457	&   22&	0.\\[-2ex]
\\[-2ex]
\# eta'(958)\\[-2ex]
331	&0.95778&	0.000203\\[-2ex]
0.445&	211&	0.13957	&-211	&0.13957	&221	&0.54751\\[-2ex]
0.294&	113&	0.769&	22&	0.\\[-2ex]
0.208&	111&	0.13498&	111&	0.13498&	221&	0.54751\\[-2ex]
0.0303&	223&	0.782&	22&	0.\\[-2ex]
0.0212	&22	&0.	&22&	0.
\end{tabular}
}
}}
\caption{\label{f:resoin} 
An excerpt from the file \texttt{resonances.input}.}
\end{figure}
\begin{itemize}
\item A record of all decay modes of one resonance starts with a line with three numbers: MC code 
(identifier) of the resonance, its mass in GeV, and its width in GeV. If --1 is put in the position 
of the mass, the code automatically reads in the mass from \texttt{pproperties[]}.
\item Record of a two-body decay contains five numbers. First, there is branching ratio for the decay 
channel, which must be multiplied by the Clebsch-Gordan coefficient, if applicable. This is followed 
by MC code of the first daughter particle and its mass, and the same for the second daughter 
particle. Again, if --1 is put instead of the mass, the mass it fed in automatically. 
\item Three-body decays are recorded with seven numbers, where MC code of the third daughter particle 
and its mass are added to the structure of the record of two-body decays. 
\end{itemize}
If the sum of branching ratios is not unity, the program will multiply them with a common factor
so that their sum will become 1. Decays into photons (or leptons) can also be included with the 
appropriate MC codes (e.g.\ 22 for photons). Any line that starts with ``\#'' or is empty is considered 
as a comment. Comments are very useful in order to enhance the readability of the records.

\subsection{Output}

Output data are directed into the file which is specified as a command line 
parameter. The possible formats for the output are explained as comments in 
\texttt{params.hpp}. Note that in case of the OSCAR1999A output format each line 
contains also two additional output numbers: On the 13th position the number of the 
fragment from which the hadron originates, or the number from which its parent 
resonance was emitted. For hadrons emitted from bulk this number is --1. 
On 14th position there is 0 if this is directly produced 
hadron or 1 if the hadrons stems from a resonance decay.

If droplet information is stored, the file will contain lines with the following 
structure: event number, number of droplet, rapidity of droplet, transverse velocity
of droplet, azimuthal angle of its transverse velocity. This is followed by the 
multiplicities of individual sorts of particles which are emitted from the given 
droplet. Thus for example a record like
\begin{quote}
\texttt{
\hspace{-1em}
\# eid   d\_id   rapidity   v\_t   phi   sorts:\\
\#   \hspace{15em}                                 -211 111 211\\
\dots \\
\# 2 \ \ 		 3		\ \ 	1.2724	\ \    0.3411\ \   2.4513 \ \   12 \ \   10\ \    16
}
\end{quote}
means that in event 2 we have droplet number 3, which moved with rapidity 1.2724
and transverse velocity 0.3411$c$ under azimuthal angle 2.4513~rad. From this droplet
12~$\pi^-$s (code -211), 10~$\pi^0$s (code 111), and 16~$\pi^+$s (code 211) are emitted. (Usually, one would 
make simulation with many more species; we only show three here for brevity.)


\section{Sample results}
\label{s:results}

\subsection{Size of droplets}

In order to get an impression of the size of droplets, we first give  
average numbers of final state hadrons coming from one droplet for various settings of parameters.
Resonances were included in this calculation. They decayed into stable hadrons and here
the final numbers of stable hadrons are given. 
They were all calculated for energy density and chemical freeze-out conditions at RHIC
$T_{ch} = 168$~MeV, $\mu_B = 266$~MeV, $\mu_S = 71$~MeV. For small droplets with the volume of 2~fm$^3$
and kinetic temperature 160~MeV, there are on average 2.3 pions, 0.2 nucleons, and 3.2 hadrons in total 
produced. For the volume 10~fm$^3$ the corresponding numbers are 8.7, 0.9, and 12.1. For very large 
volume like 100~fm$^3$ we have on average 80.9 pions, 7.8 nucleons, and 112.3 hadrons. The temperature dependence is rather weak,
e.g. in the latter case dropping the temperature from 160 to 100~MeV increases the pion number to 94. 
In general, lower kinetic temperature leads to larger number of hadrons since less energy 
is used in the form of kinetic energy.

\subsection{Comparison with THERMINATOR}

DRAGON is similar to THERMINATOR \cite{therminator},
with some differences:
\begin{itemize}
\item DRAGON allows for particle emission from fragments; if fact this was the main motivation for 
conceiving it. 
\item The radial profile of transverse expansion velocity grows linearly in DRAGON (cf. eq.~\eqref{etat}),
while it is kept at constant value if the blast wave model option is set in THERMINATOR. 
\item THERMINATOR also simulates freeze-out in Cracow single freeze-out model \cite{Broniowski:2001we}
and in blast wave model with varying time dependences of radial coordinate of the freeze-out hypersurface
\cite{Kisiel:2006is}.
\item DRAGON allows to simulate azimuthally non-symmetric fireballs. 
\item THERMINATOR uses SHARE \cite{Torrieri:2004zz,Torrieri:2006xi} to define chemical 
composition and resonance decays. The list 
of included resonances and decays is longer than that of DRAGON, which should not cause a problem,
however, as higher 
lying states are suppressed by Boltzmannian factor.  
\end{itemize}

\begin{figure}
\centerline{\includegraphics[width=1\textwidth]{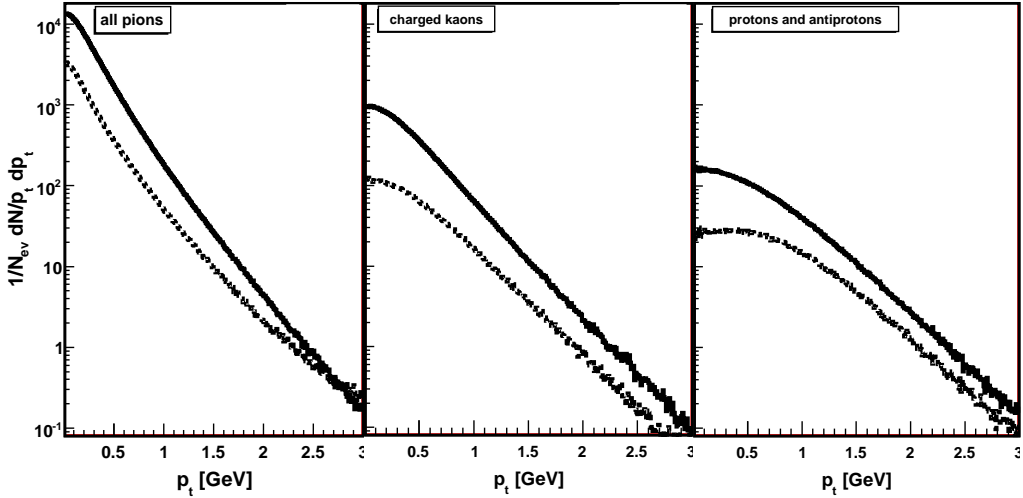}}
\caption{\label{f:THQU} 
Comparison of $p_t$ spectra generated by DRAGON (solid lines) and THERMINATOR \cite{therminator} (dotted lines).
In THERMINATOR the blast-wave source was chosen with the parameters that come with the standard distribution:
$v_r = 0.55$, $\tau = 9.74$~fm/$c$, $\rho_{\rm max} = 7.74$~fm, $T = 165.6$~MeV, $\mu_B = 28.5$~MeV, 
$\mu_S = 6.9$~MeV. Parameters were chosen correspondingly in DRAGON with $\rho_0 = 0.55$. Spectra from 
10$^4$ simulated events are shown.
}
\end{figure}

The results of both models are compared in Figure~\ref{f:THQU}.
For this comparison, THERMINATOR was run with the set of parameters with which it is distributed, just the freeze-out 
model was set to be blast wave and not the Cracow single freeze-out. Parameters of DRAGON were chosen 
correspondingly. Note that multiplicity results from the chosen parameters in THERMINATOR while it is set
by hand in DRAGON. Thus the difference in absolute normalisation of the spectra is irrelevant.
Recall, that the transverse flow expansion rapidity 0.55 was constant irrespective of 
radial coordinate in THERMINATOR; in DRAGON we chose $\rho_0 = 0.55$. This leads to the same mean transverse
expansion velocity. 

In Figure~\ref{f:THQU} we observe this difference. At low $p_t$, THERMINATOR spectra are suppressed against 
DRAGON. These are the hadrons with low transverse velocities. In DRAGON, they are produced mainly from regions 
of the fireball which move slowly outwards. In THERMINATOR, such regions are missing (since transverse velocity 
is everywhere the same) and this leads to the effect on the spectra. Pions move relativistically already at
rather low $p_t$, so for them this suppression is at \emph{very} low $p_t$ and is almost invisible. On the other 
hand, at very high transverse momentum their spectra become flatter in THERMINATOR. These are pions which move 
with \emph{very high} transverse velocities---higher than the transverse expansion velocity of the fireball. 
In THERMINATOR, their production is enhanced by the fact that we have larger region that moves outwards with 
rather large transverse velocity. In other kinematic regions---high $p_t$ for kaons and protons and semi-low
$p_t$ for pions---the spectra from DRAGON and THERMINATOR are parallel and consistent.


\subsection{Single-particle spectra}

\begin{figure}[t]
\centerline{\includegraphics[width=0.7\textwidth]{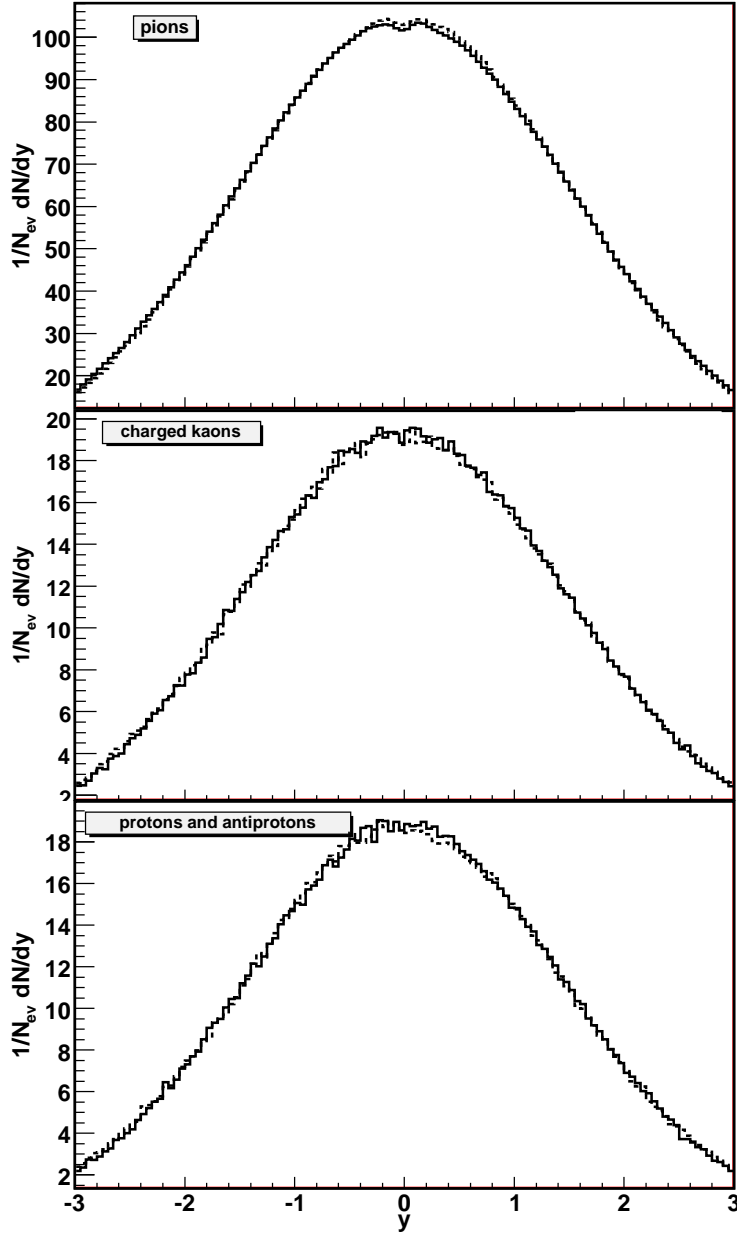}}
\caption{\label{f:yspec} 
Comparison of rapidity spectra from a model with fragments (solid lines, $b=50$~fm$^3$) and without
them (dotted lines) for charged pions (top), kaons (middle), and protons and antiprotons (bottom). 
A model with Gaussian space-time profile was chosen with the width $\Delta \eta = 1.3$. 
Other parameters were: $T_k = 160$~MeV, $T_{ch} = 168$~MeV, $\mu_B = 266$~MeV, 
$\mu_S = 71$~MeV, $\rho_0 = 0.6$, 
 Spectra from 
10$^4$ simulated events are shown.
}
\end{figure}

\begin{figure}
\centerline{\includegraphics[width=0.9\textwidth]{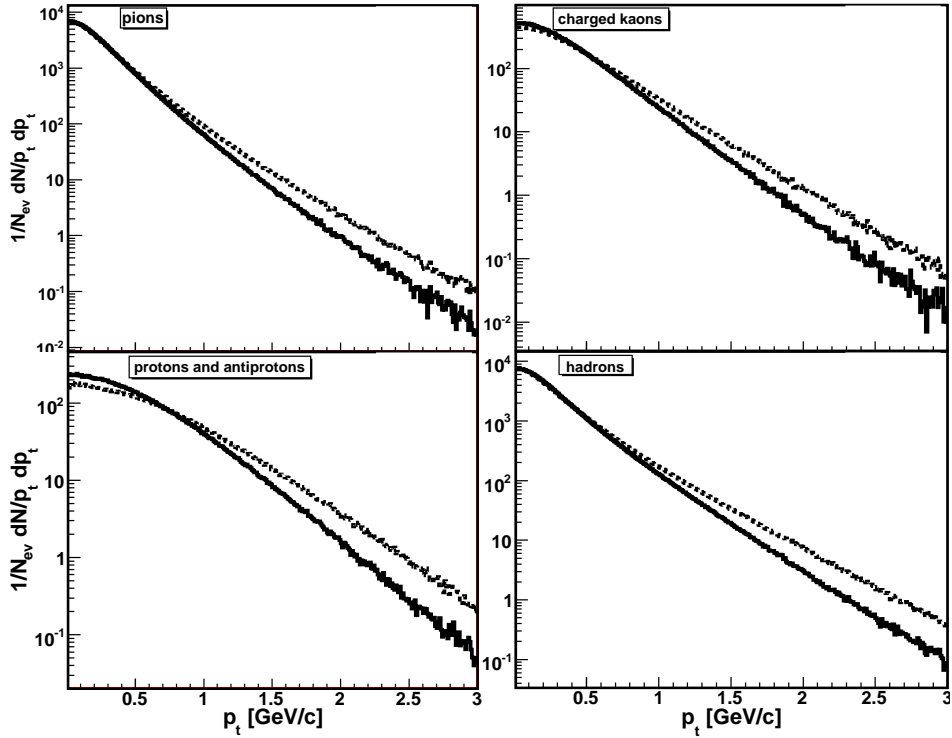}}
\caption{\label{f:sps-spec} 
The $p_t$ spectra from a model with Gaussian space-time rapidity profile with the width $\Delta \eta = 1.3$. 
Results from the same simulation as in Fig.~\ref{f:yspec}: $T_k = 160$~MeV, $T_{ch} = 168$~MeV, $\mu_B = 266$~MeV, 
$\mu_S = 71$~MeV, $\rho_0 = 0.6$. Spectra from 
10$^4$ simulated events are shown; they are integrated over rapidity. Solid lines show results from 
a simulation with fragments with $b = 50$~fm$^3$ and dotted lines show results from a simulation
without fragments. 
}
\end{figure}

\begin{figure}
\centerline{\includegraphics[width=0.9\textwidth]{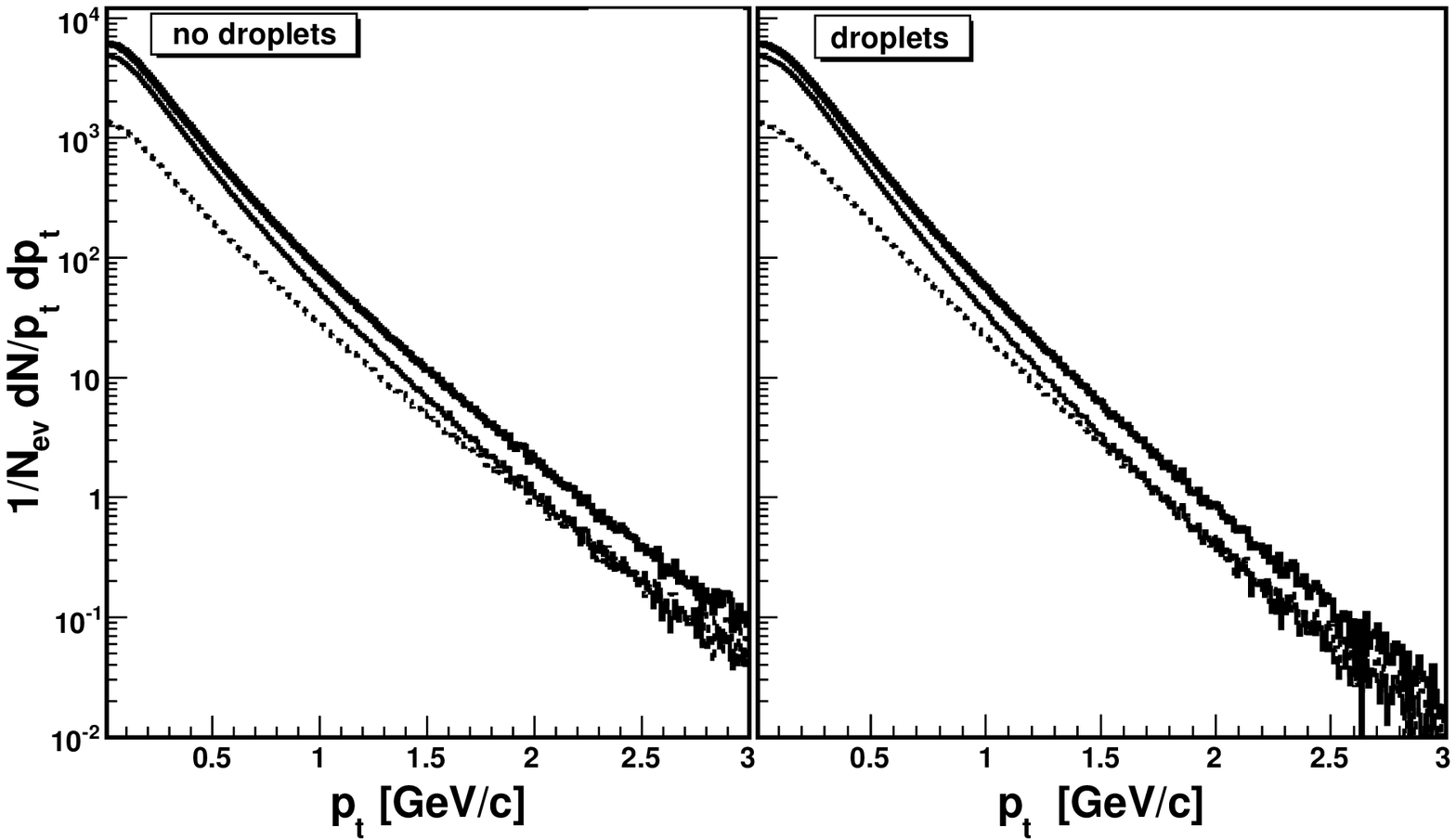}}
\caption{\label{f:anat} 
The contributions to pion spectra from the simulations of Fig.~\ref{f:sps-spec}. Thin solid 
line is the resonance contribution, dotted line shows direct pion production and thick solid line
is the total resulting spectrum. 
Left: simulation without fragments (droplets), right: simulation with fragments (droplets)
with $b = 50$~fm$^3$. 
}
\end{figure}

Results obtained from a simulation with Gaussian rapidity profile, 
are illustrated in Figures~\ref{f:yspec}, \ref{f:sps-spec}, and \ref{f:anat}. 
First of them shows the rapidity spectra. In the simulation, space-time rapidity distribution was
centered at $\eta=0$ with a width of 1.3 and kinetic freeze-out temperature 
$T_k = 160$~MeV. We show results from simulations with and without fragments. Note, that in case 
of fragments their typical size is actually chosen rather large ($b = 50$~fm$^3$). No difference is observed 
between the spectra obtained from the two kinds of simulation. This is expected: fragmentation of the 
fireball cannot be seen from observables integrated over large number of events. 

The same conclusion can be drawn from the $p_t$ spectra. The spectra from non-fragmented fireball appear 
slightly flatter. This is because in such a simulation particles can be produced from regions close to the 
radial edge of the fireball which move with highest transverse velocity. Fragments are produced in such a way 
that their volume is always completely included in the volume of the fireball and their velocity corresponds 
to the fireball expansion velocity at the \emph{centre} of the fragment. Thus fragments (especially the big ones)
can never obtain as high transverse velocity as the hadrons in non-fragmented fireball. In a case where 
both simulations are performed with the same $\rho_0$, like in Fig.~\ref{f:sps-spec}, simulation without
fragmentation thus yields slightly flatter $p_t$ spectra. 

In Fig.~\ref{f:anat} we show the contributions to pion spectra. In both cases, with and without the fragments, 
they are the same. The most important contribution at this chemical freeze-out temperature 
($T_{ch} = 168$~MeV, $\mu_B = 266$~MeV, $\mu_S = 71$~MeV) comes from resonance decays. Direct pion production 
becomes equally important at around $p_t = 1.5$~GeV/$c$.

\subsection{An illustration of fragments}

\begin{figure}
\centerline{\includegraphics[width=12cm]{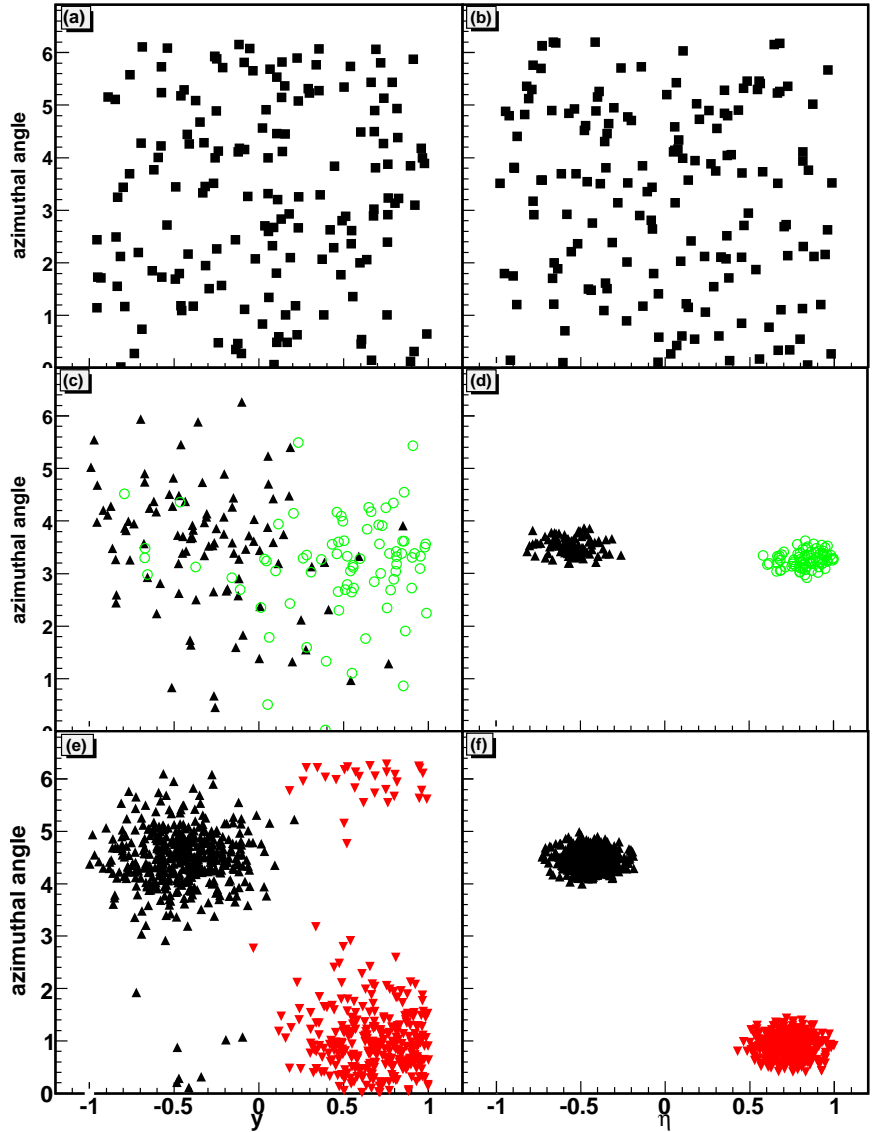}}
\caption{\label{f:drps} 
Positions in space-time rapidity $\eta$ and azimuthal angle from which pions were emitted 
(right column) as well as rapidities and azimuthal angles of the momenta which they had 
(left column). No resonances and no other particles than pions were produced here. 
Top  row: pions emitted from boost-invariantly expanding fireball which 
does not fragment. Middle row: pions emitted from two large fragments with temperature 170~MeV. 
Hadrons from the same fragment are indicated by the same symbol. Rapidities, transverse velocities,
and azimuthal angles 
of the fragments are (-0.55,~0.51$c$,~3.53~rad) and (0.84,~0.47$c$,~3.30~rad) and they emit 139 and 123 pions. 
Bottom row: as middle row but for the sake of illustration the temperature was set to 10~MeV.
Rapidities, transverse velocities and azimuthal angles of the two fragments are (-0.44,~0.45$c$,~4.49~rad) 
and (0.73,~0.34$c$,~0.92~rad) and they emit 435 and 390 pions.
}
\end{figure}

As mentioned, fragmentation cannot be observed in spectra integrated over a large number of 
events. Correlation and fluctuation observables go beyond the scope of this paper and shall be 
investigated in subsequent dedicated papers. Just to illustrate the production from fragments, 
momenta and positions of particles are illustrated in Figure~\ref{f:drps}.
Positions of emission points (right column) and momenta of hadrons (left column) in rapidity and 
azimuthal angle are shown. In this 
illustration only directly emitted pions have been considered for the sake of simplicity. In the top
row we show momenta and positions of the emission for a fireball that expands boost invariantly in 
longitudinal direction. It is observed that both positions and momenta spread uniformly over the 
whole $(\eta,\phi)$ interval. In order to clearly illustrate the effect of fragments on the momentum distribution,  
in the bottom row results from an event with large fragment volume ($b = 50$~fm$^3$) and very low
kinetic freeze-out temperature ($T_k = 10$~MeV) are shown. The large volume ensures that there are many pions
coming from one fragment and the low temperature puts  a limit on their thermal velocity so that it stays 
close to that of the fragment. Thus clustering of the emission points is transferred to the momentum 
space. Note, however, that the values of parameters are unrealistic. In order to show results in a more 
realistic simulation, in the middle row a temperature $T_k= 170$~MeV was chosen. The clustering in momentum 
space is still observed though now it is much less pronounced. Note that in realistic situation 
(i) the size of fragments will probably be smaller; (ii) the number of fragments will be bigger; (iii) also 
other hadron species shall be present and pions shall dominantly be produced from resonance decays. 
Thus clustering in momentum space shall be even less visible and sophisticated methods might be necessary 
in order to identify fragmentation.


\subsection{Elliptic flow}

\begin{figure}
\centerline{\includegraphics[width=10cm]{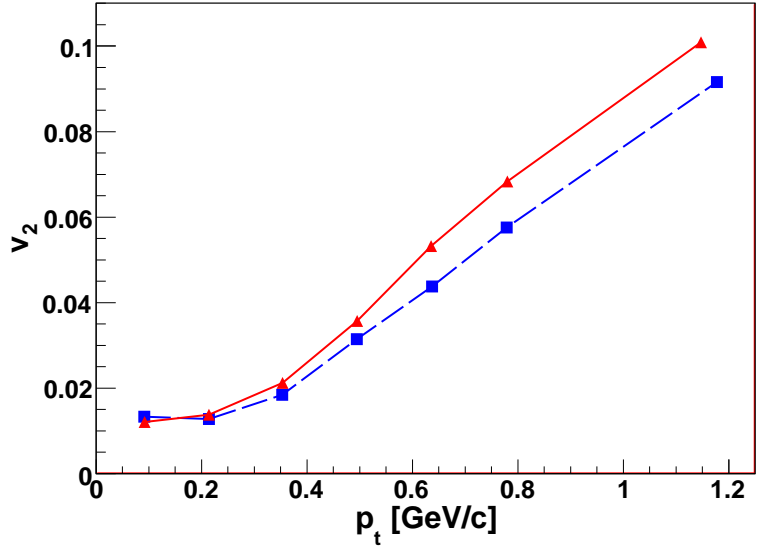}}
\caption{\label{f:vtwo} 
The elliptic flow parameter $v_2$ of all hadrons calculated for parameters of the blast wave 
model that have been fitted to Au+Au collisions at $\sqrt{s}=130\,A$GeV,
centrality class for elliptic flow 11--45\%   \cite{Adler:2001nb}:
$T_{k} = 107$~MeV, $\rho_0 = 0.85/\sqrt{2}$, $\rho_2=0.058$, $R=11.1$~fm, 
$a=0.939$, $\tau = 7.7$~fm/$c$ \cite{Retiere:2003kf} and 
$T_{ch} = 174$~MeV, $\mu_B = 46$~MeV, $\mu_S = 13.6$~MeV \cite{BraunMunzinger:2001ip}. 
Solid red lines with triangles show results from simulation with fragments ($b=10$~fm$^3$) 
and dashed blue lines with squares show results from simulation where all hadrons have 
been produced from the bulk fireball. 
}
\end{figure}

An important observable is the elliptic flow. This model provides the possibility to 
simulate azimuthally non-symmetric collisions where elliptic flow is observed. Within the blast wave
model, the existence of azimuthal anisotropy of single particle spectra can be achieved by 
setting the spatial anisotropy $a$ and/or the expansion velocity anisotropy $\rho_2$ \cite{tomasik_app}. 
In order to illustrate the elliptic flow, we calculated $v_2$ for a set of parameters which reproduce 
mid-central Au+Au collisions at $\sqrt{s}=130\,A$GeV \cite{Retiere:2003kf} (Figure~\ref{f:vtwo}). 

The elliptic flow parameter $v_2$ was determined from simulated data via two-particle correlations
in the azimuthal angle
\begin{eqnarray}
v_2(p_t) & = & \left [   \frac{\int d \Delta \, \cos \left ( 2 \Delta  \right )\,  N_2(\Delta,p_t)}%
{\int d \Delta \, N_2(\Delta,p_t)}  \right ]^{1/2}\\
N_2(\Delta,p_t) & = & \frac{1}{2}\int d\phi_1 \int d\phi_2 \, \delta(\Delta - |\phi_1 - \phi_2|)\,
N_1(\phi_1,p_t)\, N_1(\phi_2,p_t) \, ,
\end{eqnarray}
where $N_1(\phi,p_t)$ is the single-particle distribution in azimuthal angle and transverse momentum.
In practice, $v_2$ is not evaluated for a sharp value of the transverse momentum, but momenta from an interval 
in $p_t$ are taken. In this analysis, the $p_t$ range was divided into seven intervals. 

In Fig.~\ref{f:vtwo} we compare the elliptic flow for all hadrons calculated in a standard blast wave model with 
one calculated in case of all hadrons emitted from fragments. In both cases, all parameters of the model 
were the same; only the percentage of particles coming from fragmets is 0 in one and 100\% in the other case. 
A slight increase of the elliptic flow in case of fragmented fireball is observed. This is due to enhanced 
correlations of hadrons from one fragment. A thorough investigation of the elliptic flow from fragmented 
fireball shall be deferred to a dedicated paper.


\section{Conclusions}
\label{s:conc}

The new Monte Carlo generator of the final positions and momenta of hadrons represents 
the blast wave model and incorporates important additional features. Above all it is the 
possibility to simulate hadron emission from a fragmented fireball. It is possible to 
vary the planned multiplicity and the percentage of particles which are emitted from the 
fragments. Additionally, it also allows for simulation of azimuthally non-symmetric events. 
THERMINATOR, the closest related model on the market, exists only in the azimuthally symmetric version.

Thus two kinds of studies can be envisaged for  which DRAGON appears unique: (i) studies of 
correlations and fluctuations connected with fragmentation of the fireball; (ii) studies 
of spectra and correlations in non-central collisions. The main driving motivation for its
development was the former, as fragmentation may be intimately connected with the dynamics 
of the phase transition. The Monte Carlo generator can be used for designing and testing 
new observables for probing fragmentation.


\paragraph*{Acknowledgment}
I appreciate discussions with Marcus Bleicher, Andrew Jackson, Igor Mishustin, Ivan Melo, 
and Giorgio Torrieri. I thank Giorgio Torrieri for running the THERMINATOR model and for discussions 
which helped to debug the code. I also thank Michal Vaga\v{c} for the computer support.

\appendix

\section{Momentum generation from Boltzmann distribution}
\label{app:boltzmann}

In this appendix it is reviewed how the momenta of hadrons are generated 
according to Boltzmann distribution, in the rest-frame of the fragment or 
fluid element. 

First, the case $T_k=0$ is treated separately. In such a case the momentum vanishes and
the energy is put equal to the hadron mass. 

\begin{figure}
\centerline{\includegraphics[width=0.7\textwidth]{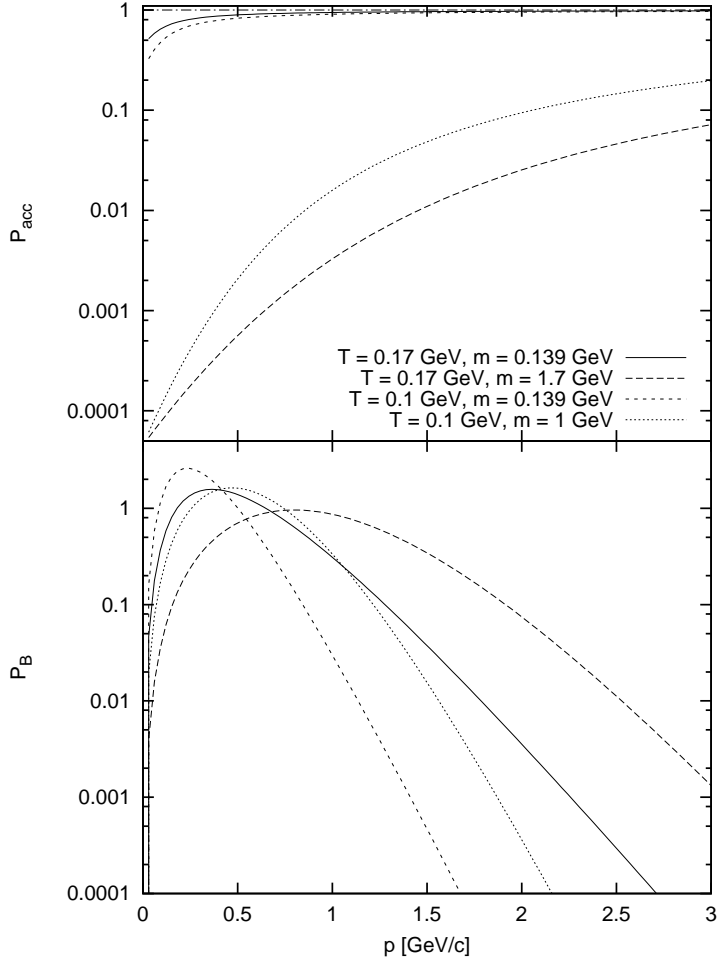}}
\caption{\label{prbs-rel} 
Top: The acceptance probability for light hadrons for some combinations of mass and 
temperature calculated from eq.~\eqref{acc-rel}. Bottom: Boltzmann distribution of momenta for light hadrons.}
\end{figure}

For non-zero temperatures, \emph{rejection method} is used to generate the size of the 
momentum. The Boltzmann distribution is 
\begin{equation}
\label{d-bol}
{\cal P}_{B}(p) \propto p^2 \exp \left ( -\frac{\sqrt{m^2 + p^2}}{T} \right )\, .
\end{equation}
If the mass of the hadron is small with respect to $T$, this distribution receives 
large contribution from the region of high momenta. Momentum is first generated according 
to gamma distribution
\begin{equation}
{\Gamma}_3(p) \propto p^{2}\, \exp\left (-\frac{p}{T}\right )\, ,
\end{equation}
which is implemented as explained in \cite{NumRec}. Once momentum $p$ is generated, it is 
accepted with probability 
\begin{equation}
\label{acc-rel}
{\cal P}_{\rm acc} = \exp\left ( \frac{p-\sqrt{m^2+p^2}}{T}\right )\, .
\end{equation}
The natural scale for the mass is the temperature. In the code, this procedure 
is used for masses smaller than $B\cdot T$, where the parameter $B$ is introduced 
as macro \texttt{BIGMASS} and is normally put equal to 10. Acceptation probabilities
of some light particles are shown in Figure~\ref{prbs-rel}.

For large masses, this procedure becomes ineffective because the acceptation 
probability is tiny if $p\ll m$. In this case,  
the problem can be treated non-relativistically. All three components of the 
momentum are generated according to the non-relativistic distribution
\begin{equation}
{\cal P}_{\rm nr,i}(p_i) \propto \exp \left ( - \frac{p_i^2}{2mT} \right )\, .
\end{equation}
Thus total momentum will satisfy the distribution 
\begin{equation}
\label{d-nonrel}
{\cal P}_{\rm nr}(p) \propto p^2 \exp \left ( - \frac{p^2}{2mT} \right )\, .
\end{equation}
This cannot be used in the rejection method in a strict way, because distribution 
\eqref{d-nonrel} drops for large $p$ as $e^{-\alpha p^2}$ whereas the correct 
relativistic distribution \eqref{d-bol} goes like $e^{-\beta p}$. Thus there 
will always be a region  of large $p$ in which $\kappa {\cal P}_{\rm nr}(p)$ becomes 
smaller than ${\cal P}_{B}(p)$ no matter the value of the multiplier $\kappa$. The acceptance 
probability calculated as ${\cal P}_{B}(p)/\kappa {\cal P}_{\rm nr}(p)$ then becomes
bigger than 1. 

\begin{figure}
\centerline{\includegraphics[width=0.7\textwidth]{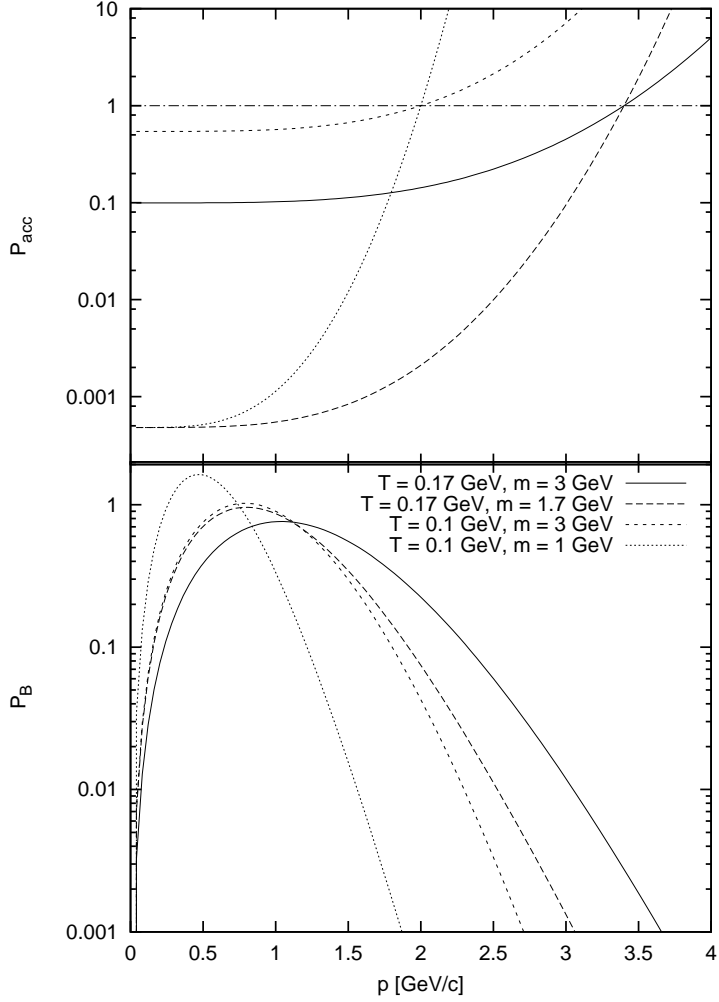}}
\caption{\label{prbs-nr} 
Top: The acceptance probability for heavy hadrons for some combinations of mass and 
temperature calculated according to eq.~\eqref{acc-nonrel}. 
Bottom: Boltzmann distribution of momenta for heavy hadrons.}
\end{figure}

In the implementation, the constant $\kappa$ is chosen in such a way that the 
acceptance probability becomes 1 at $p=AT$. Thus the acceptance probability then is
\begin{equation}
\label{acc-nonrel}
{\cal P}_{\rm acc} = \exp\left ( \frac{p^2}{2mT} -\frac{\sqrt{m^2+p^2}}{T}\right )
\exp\left ( - \frac{A^2 T}{2m} + \sqrt{\frac{m^2}{T^2} + A^2} \right )\, .
\end{equation}
This technically means that for $p$ larger than $AT$ all hadrons 
with momenta generated according to non-relativistic distribution are accepted. It also means
that in the spectrum these are suppressed relative to the case where they would be 
generated according to the correct distribution, due to faster decrease of the non-relativistic
distribution. The parameter $A$ must be chosen. If it is small, then as a result the algorithm 
will produce too few hadrons with momenta above $AT$. On the other hand, if $A$ is too large, then 
the acceptance probability will become very small at small momenta and the algorithm becomes 
ineffective or even not working. For usual freeze-out temperatures, between say 100 and 180 MeV, 
the default reasonable setting is $A=20$. For 
this choice, in case $T = 0.1$~GeV and $m = 3\, \mbox{GeV}/c$ the fraction of hadrons which is 
suppressed due to this artifact is at the level 0.86\%. Increasing temperature and decreasing 
the mass makes this fraction yet smaller. On the other hand if this algorithm is used
for masses larger than $10T$, then minimum acceptance probability is reached for the smallest
mass $10T$ and $p=0$ and is about $4.8\cdot 10^{-4}$. 
In Figure~\ref{prbs-nr} the acceptance 
probabilities as functions of momentum for heavy (non-relativistic) hadrons are shown.

\end{document}